# Hotspot Prevention Through Runtime Reconfiguration in Network-On-Chip *

G. M. Link, N. Vijaykrishnan

*The Pennsylvania State University, University Park, PA, 16802*

{*link,vijay*}*@cse.psu.edu*

## Abstract

*Many existing thermal management techniques focus on reducing the overall power consumption of the chip, and do not address location-specific temperature problems referred to as hotspots. We propose the use of dynamic runtime reconfiguration to shift the hotspot-inducing computation periodically and make the thermal profile more uniform. Our analysis shows that dynamic reconfiguration is an effective technique in reducing hotspots for NoCs.*

## 1. Introduction

Technology scaling and the quest for higher performance has caused the power density of chips to reach alarming levels. The temperature profiles are spatially non-uniform across the chip and vary based on the power dissipation characteristics of the individual chip components, creating localized hot spots [1], resulting in failures. These hot spots are not effectively addressed by techniques that minimize the power consumption of the chip. Thermal solutions employed in current commercial processors such as dynamic clock disabling and dynamic frequency scaling stop or shut down the entire chip for brief periods of time. Instead of shutting down or slowing down the entire chip, recent proposals have focused on the migration of the workload from a hot component to a cooler spare until the temperature reduces [2]. However, the use of spare units adds to hardware redundancy consuming additional area. In contrast to using spares, our work explores the use of dynamic reconfiguration as a mechanism for addressing localized hotspots. We periodically modify the configuration data stored to migrate the functionality across the chip in order to balance the temperature profile. In essence, we implement a spatial remap of the different functionalities at runtime. Our solution can be employed in programmable embedded architectures such as Network-on-Chip (NoC) designs and Field Programmable devices that are being increasingly used in various applications.

We evaluate our technique as applied to a Low Density Parity Check (LDPC) Decoder implemented on NoC with a thermally-aware static mapping. We explore issues such as the frequency of functional migration, choice of remap function, techniques for modifying of the configuration stream, performance impact of the reconfigured mappings, and heating patterns that can(not) be effectively handled by some migration schemes. Our experimental investigation reveals that dynamic reconfiguration approaches are successful in balancing the thermal profiles and in reducing the peak temperature by up to 8° C (from 85C) when starting with a thermally optimized static mapping.

* This work was supported by a MARCO/DARPA PAS grant, as well as NSF 0093086.

## 2. Experimental Platform

Our experimental platform is based on the HotSpot thermal library. The HotSpot tool was left with all settings at the default values and an ambient temp. of 40° C.

Our floorplans were taken directly from the layout of our sample chips. Two test chips implementing LDPC decoding [3] were synthesized and placed and routed using a commercial 160nm standard cell library. Overall, each functional unit has an area of 4.36 sq. mm. Power consumption of each unit is determined through the use of Synopsys Power Compiler. A modified cycle-accurate NoC simulator is then run with an encoded message to obtain switching rates for the components in the chip during operation. Note that our simulations also include the energy consumed during the migration operation to more accurately evaluate the utility of our proposed method.

To more fully explore the design space for reconfiguration, we evaluate multiple configurations of our test chips. In particular, the 4x4 chip is evaluated with two different configurations (referred to as A and B), while the 5x5 chip is evaluated with three different configurations (C, D, E).

Differences in thermal profiles and power consumption between the configurations are due to the irregularity of the communication patterns and the amount of computation mapped to a single PE. Consequently, our workload was mapped onto PEs using a thermally-aware placement algorithm that minimizes the peak temperature. Using such a thermally-aware mapping puts our method in a worst-case light, that where power and heat distribution has already been equalized at design time.

### 2.1. Techniques for Implementing Remapping

As maintaining additional configuration registers on chip for remapping purposes requires a significant increase in chip area, we propose a means of transforming existing configuration information at run-time to generate new placements. In this scheme, the operation of the PEs is halted, the configuration and state information of each PE is passed through a conversion unit, and then sent across the network to the destination PE.

### 2.2. Migration Schemes

Our avenue of investigation is based on a logical model that ensures the new position of the workloads can be algebraically determined from the current position information and that the workloads will retain the same relative position to each other, resulting in a much more predictable impact on network traffic patterns. Intuitively, we can see that there are only a limited number of ways of placing the logical functionality onto the chip. If we abstract this relative positioning requirement into a theoretical plane in which all workloads are statically placed, we can see that all possible migrations must operate on the plane as a whole, rather than on the workloads themselves. In practice, there are a total of three ways of adjusting a plane that, when combined, define all possible operations. These three operations are rotation, mirroring, and translational shifting. We consider each of these three operations as migration functions separately. During the migration operation, it is possible to ensure congestion-free packet movement by transforming groups of PEs in phases. This congestion-free operation allows for deterministic migration times, making our technique applicable to real-time systems.

| | New X Coordinate | New Y Coordinate |
|---|---|---|
| Rotation | N-1-Y | X |
| X Mirroring | N-1-X | Y |
| X Translation | X + Offset | Y |

**Table 1. Transformation Functions**

### 2.3. Implementation of the Migration Functions

All of the proposed migration functions are mathematically quite simple, and require little hardware to properly implement. Each migration function takes as input the current X, Y location of the workload, and provides as output the new X, Y destination of the workload. As such, only 3-bit operands are required to address up to 64 PEs, resulting in fast operation. The transformation functions are extremely simple when represented in {X,Y} format, and are shown in Table 1. The selection of these simple functions allows the migration unit to remain small, fast, and low power. More importantly, the simplicity and predictability of the migration functions presented allows for a simplified I/O interface to the outside of the chip, by transforming the destination address assigned to all incoming packets and transforming the source address of all packets leaving the chip. By including a migration unit at the I/O interface, the migration operation is totally transparent to the outside world. Finally, we note that the same migration unit can perform all migration functions presented with only minor changes to the mathematical operations, allowing dynamic alteration of the migration function at runtime

### 3. Results

We begin our analysis by comparing the relative effectiveness of the different migrations at reducing the peak temperature of the test chips. Figure 1 shows the reduction in peak temperature in the various circuit configurations with different migration techniques. For circuit configurations A and B, the rotational and X-Y mirroring migrations reduce the peak temperature the most, while for the larger configurations, translation is more effective. This difference in efficacy is due, in large part, to the even dimensionality (4x4 array) of test cases A and B, as opposed to the odd dimensionality (5x5 array) of test cases C, D, and E. In the odd-dimensioned test cases, both the rotational and mirroring migration functions ignore the central PE, and as such, they are unable to balance the heat generated at the center of the device. The poor behavior of the right shift is due to the relative power output of the rows in the various test cases. In all test cases, one of the rows had a significantly higher power output than the remaining rows, generating a warm band that right shifting alone is unable to distribute. While such a warm band might seem to skew our results, note that a thermally-aware placement algorithm was used to generate our initial test cases, and as such, it is reasonable to assume that such characteristics would be even more common in non-thermally-aware placements.

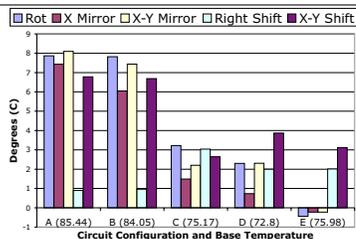

**Figure 1. Reduction in Peak Temps**

Among the configurations tested, X-Y shifting has the highest average temperature reduction, 4.62° C. Rotational migration has the second highest average temperature reduction, 4.15° C, but actually results in higher peak temperatures for configuration E for two reasons. The first reason, common to both the rotational and mirroring migrations, is that the hotspots in configuration E are near the center of the chip, where those algorithms are least efficient at migrating workload away from the hotspot. Second, the rotational migration has the largest energy penalty for performing reconfiguration, resulting an increase in average chip temperature of 0.3° C. All of the above simulations were performed with a migration period of 109 microseconds, resulting in an overall throughput reduction of 1.6%. Higher frequencies of migration result in a more even temperature distribution, but do so at the cost of performance. For a reconfiguration period of 437.2 microseconds, the overall performance penalty drops to less than 0.4%, and the peak temperatures rise less than a tenth of a degree in the additional time between migrations. Further, we can increase the period between reconfigurations to 874.4 microseconds and reduce the throughput penalty to less than 0.2% without significant impact on peak temperature. We note that the periods for reconfiguration were chosen to coincide with the completion of the decoding of LDPC message blocks, minimizing the amount of state information that must be transferred between PEs. The overhead for this state information was included in our simulations.

### 4. Conclusions and Future Work

In this paper, we propose the use of dynamic runtime reconfiguration to shift the hotspot-inducing computations periodically and make the thermal profile more uniform. Different approaches to reconfiguration are proposed and evaluated for their effectiveness using a target Network on Chip designed in 160nm technology to implement a LDPC decoder. Our analysis shows that dynamic reconfiguration reduces the peak temperature up to 8° C over a thermally optimized static placement. Our work demonstrates that hotspots can be a problem even in homogenous architectures, such as FPGAs, and that workload migration is an effective technique for reducing hotspots, even when thermally-aware placement is used at design time.